\documentclass[conference]{IEEEtran}
\IEEEoverridecommandlockouts

\usepackage{cite}
\usepackage{amsmath,amssymb,amsfonts}
\usepackage{graphicx}
\usepackage{booktabs}
\usepackage{multirow}
\usepackage{url}
\usepackage[hidelinks]{hyperref}
\usepackage{xcolor}

\begin{document}

\title{GestaltMML: Enhancing Rare Genetic Disease Diagnosis through Multimodal Machine Learning Combining Facial Images and Clinical Text}
\author{
\IEEEauthorblockN{Da Wu\textsuperscript{1}, Zhanliang Wang\textsuperscript{1,2}, Hongzhuo Chen\textsuperscript{1,2}, Jingye Yang\textsuperscript{1}, Cong Liu\textsuperscript{3}, Tzung-Chien Hsieh\textsuperscript{4},\\
Elaine Marchi\textsuperscript{5}, Justin Blair\textsuperscript{6}, Peter Krawitz\textsuperscript{4}, Chunhua Weng\textsuperscript{3}, Wendy Chung\textsuperscript{7}, Gholson J. Lyon\textsuperscript{5,8},\\
Ian D. Krantz\textsuperscript{6}, Jennifer M. Kalish\textsuperscript{6,9,10}, Kai Wang\textsuperscript{1,11,*}}
\IEEEauthorblockA{\footnotesize
\textsuperscript{1}Raymond G. Perelman Center for Cellular and Molecular Therapeutics, Children's Hospital of Philadelphia, Philadelphia, PA 19104, USA\\
\textsuperscript{2}Department of Mathematics, University of Pennsylvania, Philadelphia, PA 19104, USA\\
\textsuperscript{3}Department of Biomedical Informatics, Columbia University Irving Medical Center, New York, NY 10032, USA\\
\textsuperscript{4}Institute for Genomic Statistics and Bioinformatics, University Hospital Bonn, Rheinische Friedrich-Wilhelms-Universit\"at Bonn, Bonn, Germany\\
\textsuperscript{5}Department of Human Genetics, New York State Institute for Basic Research in Developmental Disabilities, Staten Island, NY, USA\\
\textsuperscript{6}Division of Human Genetics, Children's Hospital of Philadelphia, Philadelphia, PA 19104, USA\\
\textsuperscript{7}Department of Pediatrics, Boston Children's Hospital, Harvard Medical School, Boston, MA, USA\\
\textsuperscript{8}Biology PhD Program, The Graduate Center, The City University of New York, New York, NY, USA\\
\textsuperscript{9}Department of Genetics, Perelman School of Medicine, University of Pennsylvania, Philadelphia, PA, USA\\
\textsuperscript{10}Department of Pediatrics, Perelman School of Medicine, University of Pennsylvania, Philadelphia, PA, USA\\
\textsuperscript{11}Department of Pathology and Laboratory Medicine, Perelman School of Medicine, University of Pennsylvania, Philadelphia, PA 19104, USA\\
\textsuperscript{*}Corresponding author: wangk@chop.edu}
}

\maketitle

\begin{abstract}
Individuals with suspected rare genetic disorders often undergo multiple clinical evaluations, imaging studies, laboratory tests, and genetic tests over a prolonged period of time, a process commonly described as the diagnostic odyssey. Addressing this odyssey has substantial clinical, psychosocial, and economic benefits. Many rare genetic diseases have distinctive facial features that artificial intelligence algorithms can use to facilitate clinical diagnosis, to prioritize candidate diseases for further laboratory or genetic testing, and to support the phenotype-driven reinterpretation of genome or exome sequencing data. Existing methods that use frontal facial photographs were built on conventional convolutional neural networks, rely exclusively on facial images, and cannot capture non-facial phenotypic traits or demographic information that are essential for accurate diagnosis. Here we introduce GestaltMML, a multimodal machine learning approach based solely on the Transformer architecture. It integrates facial images, demographic information (age, sex, ethnicity), and clinical notes (optionally a list of Human Phenotype Ontology terms) to improve prediction accuracy. We evaluate GestaltMML on 528 diseases from the GestaltMatcher Database and on several in-house and published cohorts, including Beckwith-Wiedemann syndrome, Sotos syndrome, NAA10-related neurodevelopmental syndrome, Cornelia de Lange syndrome, and KBG syndrome. GestaltMML improves on the state-of-the-art image-only ensembled model, narrows the diagnostic accuracy gap for patients from under-represented ancestries, and clarifies when multimodal fusion is beneficial and when image-only inference is preferable. The results suggest that GestaltMML can greatly narrow the candidate diagnoses of rare diseases and may facilitate the reinterpretation of sequencing data.
\end{abstract}


%==============================================================
\section{Introduction}
%==============================================================
A substantial proportion of the global population, more than 6\%, is affected by a rare genetic disorder \cite{r1}. While collectively common, rare diseases are individually rare \cite{r2}. They are typically defined as affecting fewer than 200{,}000 people in the USA or less than one in 2{,}000 of the general population in Europe \cite{r3}. Based on the latest Orphanet \cite{r4} and OMIM \cite{r5} databases, there are at least 7{,}000 rare genetic diseases. Because of the inherent rarity and the extensive phenotypic heterogeneity, making an accurate genetic diagnosis is a challenge that often leads to a diagnostic odyssey \cite{r6,r7,r8}. Patients with suspected genetic syndromes frequently undergo multiple clinical evaluations, imaging studies, and laboratory tests, in addition to genetic tests such as karyotype, chromosome microarray, gene panels, exome sequencing, or genome sequencing. Clinicians often struggle to decide which diagnostic test to use, because they must navigate long differential diagnoses for many different symptoms. Shortening the odyssey could have significant clinical, psychosocial, and economic benefits \cite{r8,r9}.

Many genetic diseases have distinctive facial features or dysmorphism, collectively considered the facial gestalt, which provide important clues for diagnosis and can expedite referrals to domain experts or suggest targeted genetic tests. In some cases the recognition of a syndrome from a facial gestalt can be the first step in making a diagnosis \cite{r10}. However, the effectiveness of facial recognition relies heavily on the clinician's experience. Given the many hundreds of rare genetic diseases with facial dysmorphisms, some identified only in the last five years, the facial recognition task is prohibitive for any clinician.

Following recent success in computer vision, several next-generation phenotyping (NGP) approaches have been developed to analyze and predict rare genetic disorders from 2D frontal facial images \cite{r11,r12,r13}. One widely known approach is DeepGestalt \cite{r11}, developed by FDNA as the Face2Gene product, which was pretrained on a deep convolutional neural network using CASIA \cite{r14} and later fine-tuned on more than 17{,}000 patient facial images covering 216 disorders. DeepGestalt was trained on a limited number of syndromes, and adding a newly discovered syndrome requires collecting new images and retraining the model. To make the model more inclusive for new unseen syndromes, GestaltMatcher \cite{r12} was introduced, which takes the feature layer before the final classification layer as a common embedding space, also known as the Clinical Face Phenotype Space, to encode learned facial dysmorphic features. Every frontal facial image is encoded into a 320-dimensional feature vector, which allows the distance between images to be quantified and the closest match among patients with known or unknown disorders to be identified, regardless of prevalence. GestaltMatcher also avoids altering the model architecture or retraining when integrating newly identified syndromes. Despite these successes, both DeepGestalt and GestaltMatcher use a relatively dated architecture and transfer-learning datasets \cite{r14}. More recently, Hustinx et al. \cite{r13} updated the architecture with iResNet \cite{r15} and ArcFace \cite{r16} and used updated facial image datasets, including VGG2 \cite{r17}, CASIA \cite{r14}, MS1MV2 \cite{r16}, MS1MV3 \cite{r16}, and Glint360K \cite{r18}, for pretraining. They also tried different loss functions and proposed a model ensemble combining three ArcFace models to integrate face verification and disorder-specific models. This ensembled image model achieves higher accuracy on unseen syndromes than all previous models after fine-tuning.

Nonetheless, facial images alone often do not provide adequate information for a precise diagnosis. Syndromes such as Noonan, Prader-Willi, Silver-Russell, and Aarskog-Scott all have severe to moderate short stature \cite{r19} that cannot be reflected in frontal facial pictures. Additional phenotypic traits such as sleep disturbances, impaired balance, and intellectual disability cannot be captured by facial or body photos and require other data types such as clinical notes. Moreover, numerous investigations \cite{r20,r21,r22,r23,r24,r25,r26,r27,r28} have examined the contribution of age, sex, and racial and ethnic differences to the phenotypic expression and frequency of various disorders. Groups often categorized as minorities encounter challenges that stem from systemic biases in data availability, collection, and analysis, which can lead to misrepresentations and disparities in rare disorder predictions for these groups. Motivated by these facts, some models already try to integrate facial images and clinical HPO terms. PEDIA \cite{r29} incorporates sequence variant interpretation with insights from DeepGestalt, combining expert human evaluation and artificial intelligence to provide a more comprehensive assessment. More recently, PhenoScore \cite{r30} was introduced with two modules, facial feature extraction from 2D photographs and HPO-based phenotypic similarity, and uses a trained support vector machine for classification. However, these models process images and texts separately and then combine the results. This type of late-fusion approach may lose information, because it fails to capture the interaction between modalities during training and uses ad hoc methods to assign weights and combine information. A related development is DxGPT \cite{r31}, a text-only GPT-based model for rare-disease diagnosis built on the closed-source GPT-4. In light of this, our objective is to create a multimodal machine learning methodology with a sophisticated modality interaction module that handles both facial images and clinical texts in a uniform manner, merging patient facial images with demographic details and clinical notes while preserving the integrity and richness of the data.

Recent progress in Transformer-based multimodal machine learning makes this objective attainable. The story of Transformers started with the landmark paper Attention Is All You Need \cite{r32}, which introduced self-attention and enabled models to process sequences in parallel rather than sequentially. Since then, Transformers have been applied across natural language processing and computer vision \cite{r33,r34,r35,r36,r37,r38,r39,r40,r41}, and several multimodal models have leveraged them, including ViLT \cite{r41}, CLIP \cite{r42}, VisualBERT \cite{r43}, ALBEF \cite{r44}, and Gemini \cite{r45}. For the task of predicting rare genetic disorders, we introduce GestaltMML, which uses ViLT \cite{r41}, the simplest vision-and-language Transformer, since it uses a non-trivial Transformer module for modality interaction while keeping the visual and textual embeddings linear. In this work we show that (i) early multimodal fusion in GestaltMML outperforms the state-of-the-art image-only ensembled model on the GestaltMatcher Database; (ii) a modality-masking analysis quantifies the contribution of images and texts; (iii) demographic and clinical text reduce the diagnostic accuracy gap for under-represented ancestries; and (iv) validation on five external cohorts yields a practical rule for when multimodal fusion helps and when image-only inference is preferable.

%==============================================================
\section{Related Work}
%==============================================================
\subsection{Image-Only Next-Generation Phenotyping}
DeepGestalt \cite{r11} is a convolutional network fine-tuned on patient photographs over a fixed set of 216 disorders, so adding a new syndrome requires retraining. GestaltMatcher \cite{r12} removed this constraint by encoding each face into a 320-dimensional descriptor in a shared Clinical Face Phenotype Space, turning diagnosis into a nearest-neighbor search that admits new disorders without retraining. Hustinx et al. \cite{r13} modernized the backbone with iResNet \cite{r15} and an ArcFace \cite{r16} objective pretrained on large face corpora \cite{r14,r17,r18}, and ensembled three models to reach the strongest reported accuracy on both seen and unseen syndromes; this ensembled image model is our image-only baseline. All three are convolutional and image-only, so they cannot use non-facial phenotypes or demographic context.

\subsection{Models Combining Images and Texts}
PEDIA \cite{r29} fuses DeepGestalt image scores with variant-level evidence for phenotype-driven exome reinterpretation, and PhenoScore \cite{r30} feeds facial features and HPO-based similarity into a support vector machine. Both are late-fusion: each modality is scored independently and merged with a hand-chosen rule, so cross-modal interactions are never learned and the combination is brittle when a modality is missing or noisy. Text-only models such as DxGPT \cite{r31} and recent chain-of-thought/retrieval-augmented LLM pipelines that diagnose directly from clinical notes \cite{r82} discard the facial gestalt entirely.

\subsection{Multimodal Transformers}
Vision-and-language Transformers instead learn cross-modal interactions jointly, differing mainly in the image embedder: VisualBERT \cite{r43} uses detector region features, CLIP \cite{r42} contrastively aligns whole-image and whole-text embeddings, ALBEF \cite{r44} aligns before fusing, and Gemini \cite{r45} scales to many modalities. ViLT \cite{r41} is the simplest, replacing the heavy visual embedder with a single linear patch projection so that capacity concentrates in the shared interaction Transformer. We build on ViLT for exactly this reason: it fuses images and text early while staying light enough to fine-tune on scarce rare-disease data. GestaltMML is, to our knowledge, the first convolution-free Transformer to perform early fusion of facial images, demographics, and clinical text for rare-disease classification.

%==============================================================
\section{Methods}
%==============================================================
\subsection{Overview and Data Sources}
\begin{figure}[t]
\centering
\includegraphics[width=\columnwidth]{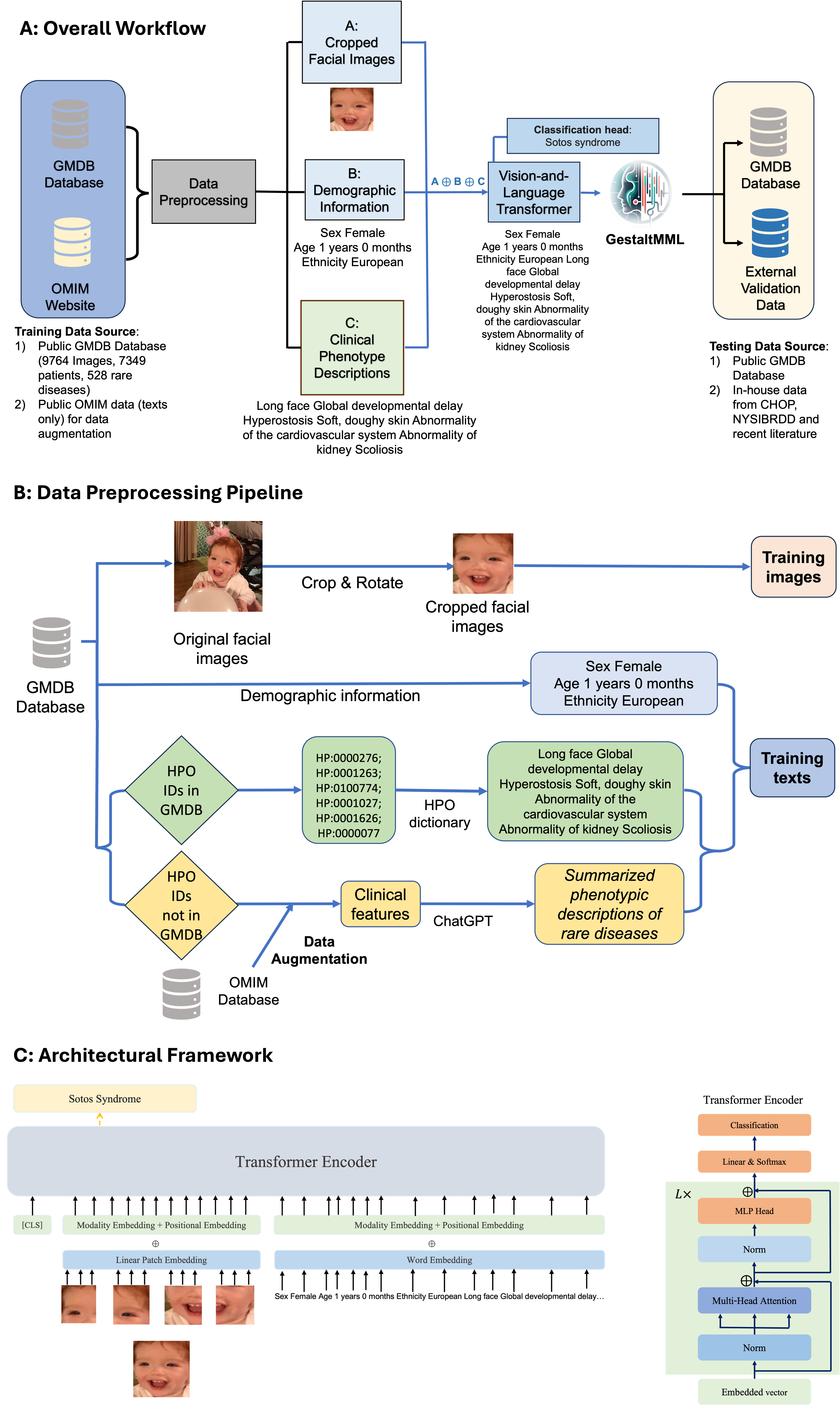}
\caption{Overview of GestaltMML. (A) Overall workflow: the model uses facial images after preprocessing, demographic information, and clinical phenotype descriptions from GMDB (when available) and from OMIM, and predicts a ranked list of candidate disorders. (B) Data preprocessing pipeline, using Sotos syndrome as an example: facial images are cropped to $112 \times 112$, and training texts are of two kinds, demographic information combined with HPO terms, and demographic information combined with OMIM clinical features summarized by a large language model. (C) Architecture: based on ViLT, GestaltMML uses a Transformer encoder over the concatenated image and text tokens.}
\label{fig:overview}
\end{figure}

The overall workflow and architecture are summarized in Fig.~\ref{fig:overview}. GestaltMML uses two data sources. The first is the GestaltMatcher Database (GMDB) \cite{r12,r46}, a curated collection of frontal facial images of genetic syndromes with corresponding textual metadata, which is open to medical researchers on application. The version used here (v1.0.9) contains 9{,}764 frontal facial images from 7{,}349 patients across 528 rare genetic disorders (Table~\ref{tab:data}). The database includes patients of diverse ancestry, but the distribution is significantly skewed, with 59.48\% of patients of European ancestry, and the age distribution is uneven, with 64.90\% of patients under five years old (see Appendix~\ref{app:dist}). This imbalance introduces difficulties for AI models and motivated us to include ethnicity in the textual data. The second source is the OMIM website \cite{r5}, which serves as a knowledge base for rare genetic disorders. To deal with the large amount of missing textual data in GMDB, we use the textual data from OMIM for data augmentation~\cite{r72}.

\begin{table}[t]
\caption{Overview of the GMDB (v1.0.9) dataset. GMDB-frequent contains disorders with more than six patients; GMDB-rare contains disorders with six or fewer.}
\label{tab:data}
\centering
\renewcommand{\arraystretch}{1.2}
\begin{tabular}{lrrrr}
\toprule
\textbf{Subset} & \textbf{Images} & \textbf{Patients} & \textbf{Disorders} & \shortstack{\textbf{Images}\\\textbf{w/ HPO}} \\
\midrule
GMDB-frequent & 8{,}547 & 6{,}376 & 244 & 3{,}962\\
GMDB-rare     & 1{,}217 & \phantom{0,}973 & 284 & \phantom{0,}470\\
\midrule
\textbf{Total} & \textbf{9{,}764} & \textbf{7{,}349} & \textbf{528} & \textbf{4{,}432}\\
\bottomrule
\end{tabular}
\end{table}

\subsection{Image Preprocessing}
Training and test images were cropped using the open-source FaceCropper described in \cite{r13} to a size of $112 \times 112$. Alternatively, an image can be resized manually so that the primary face fills the picture. The original facial images are only cropped and rotated, without flipping or conversion to grayscale. We deliberately avoid these common augmentations because facial dysmorphism is often asymmetric or side-specific, and operations such as horizontal flipping or color removal could erase exactly the subtle cues that distinguish one syndrome from another.

\subsection{Text Preprocessing}
We separate text preprocessing into two cases. The first case concerns images that have at least one HPO identifier in the present-features column of the metadata. Here we transform the HPO identifiers into text using the standard HPO dictionary \cite{r74} and concatenate them with spaces. For example, ``HP:0000486; HP:0001263; HP:0010864'' becomes ``Strabismus Global developmental delay Intellectual disability severe.'' We then prepend the patient demographics, which contain sex, age, and ethnicity, for example ``Sex male Age 4 years 8 months Ethnicity European,'' leaving missing fields blank. The combined text then reads ``Sex male Age 4 years 8 months Ethnicity European Strabismus Global developmental delay Intellectual disability severe.''

The second case concerns images that have no present features. Here we use the clinical-features section of OMIM as the primary source for data augmentation. Because of input-length limits and to save computing resources, we summarize these texts to within 500 tokens using OpenAI's ChatGPT \cite{r75}, with the prompt ``Summarize the most crucial phenotype characteristics of the following texts describing clinical features of some rare genetic disorder within 500 tokens.'' A sample summarized paragraph reads ``Clinical features of this rare genetic disorder include supravalvular aortic stenosis SVAS mental retardation distinctive facial features dental anomalies peripheral pulmonary artery stenosis infantile hypercalcemia statural deficiency characteristic dental malformation and a hoarse voice.'' We remove commas, periods, colons, and parentheses to save token space, and again prepend demographics. This preprocessing is tailored to GMDB. For clinical practitioners in real-world settings, we recommend using PhenoGPT \cite{r73} to extract HPO terms from clinical text paragraphs and then concatenating them with demographic data as described.

\subsection{Model Architecture}
GestaltMML is built on ViLT \cite{r41}, which keeps the per-modality embedding stage deliberately light and concentrates capacity in a shared Transformer encoder. The architecture is summarized in Fig.~\ref{fig:overview}C and proceeds as follows.

\emph{Image embedding.} The cropped $112 \times 112$ facial image is divided into a grid of non-overlapping patches. Each patch is flattened and mapped to a token by a single linear projection, the same scheme used by ViT \cite{r37}. Unlike convolutional or region-based embedders, this stage has very few parameters and performs no feature extraction of its own, so the burden of learning facial structure falls on the shared encoder.

\emph{Text embedding.} The concatenated demographic and HPO string is tokenized and mapped to word-embedding vectors in the standard manner for Transformer language encoders. A special classification token is prepended to the sequence and its final representation is used for prediction.

\emph{Fusion.} The image and text tokens are concatenated into a single sequence. To let the encoder distinguish the two streams and preserve order, each token receives a modal-type embedding and a positional embedding in addition to its content embedding. The combined sequence is then processed by a stack of identical Transformer encoder blocks, each consisting of multi-head self-attention, a position-wise feed-forward network, layer normalization, and residual connections. Because self-attention operates over the full concatenated sequence, every layer can relate any image patch to any phenotype token. This is the essence of early fusion: the two modalities interact throughout the network rather than being scored separately and combined at the end, which is the key architectural difference from PEDIA \cite{r29} and PhenoScore \cite{r30}.

\emph{Prediction head.} The final representation of the classification token is passed through a multilayer perceptron head followed by a softmax over the 528 disorders, yielding a ranked list of candidate diagnoses. The model is entirely convolution-free, consistent with the principle that attention can replace recurrence and convolution \cite{r32,r37}.

\subsection{Training}
We fine-tune ViLT \cite{r41}, which was pretrained on MSCOCO \cite{r76}, Visual Genome \cite{r77}, SBU Captions \cite{r78}, and Conceptual Captions \cite{r79}, and then evaluate on the various test sets, which include both GMDB and external validation data. Fine-tuning uses a cross-entropy objective over the disorder classes, and the optimization settings follow the publicly released configuration in our code repository. To make the results robust to the particular train-test partition, every experiment is repeated with three random seeds, and we report the mean and standard deviation of each accuracy metric. The OMIM-based text augmentation is applied only during training, so that images lacking curated HPO terms can still contribute a meaningful textual signal; this is important because a large fraction of GMDB images have no recorded present features (Table~\ref{tab:data}).

\subsection{Baselines and Modality-Masking Variants}
Our image-only baseline is the ensembled image model \cite{r13}, the state-of-the-art iResNet and ArcFace model ensemble, which we denote Ensembled. To study the contribution of each modality within our own architecture, we define two masked variants. GestaltViT is fine-tuned with the entire text component replaced by a placeholder, isolating the image signal, and is evaluated only on images with existing HPO terms. GestaltLT fixes a single patient photo across all training samples, isolating the text signal, with all other aspects unchanged. The name LT stands for Language Transformer. All variants are fine-tuned on GMDB (v1.0.9).

\subsection{Optimal Train-Test Splits}
\label{sec:optimal_splits}
Because much of the text is missing, we construct optimal train-test splits in which the test set contains only images that have non-null present features and whose disorders also appear with features in the training set. For a train-test ratio of $x{:}1$, let $D$ be the set of disorders with such images, and for a disorder $d$ let $S_d$ be the set of qualifying image identifiers. We randomly assign $\lfloor |S_d|/(x{+}1)\rfloor$ of them to the test set:
\begin{equation}
T = \bigcup_{d \in D} S_d^{\text{test}}, \qquad
\big|S_d^{\text{test}}\big| = \Big\lfloor \tfrac{|S_d|}{x+1} \Big\rfloor .
\end{equation}
For example, with $|S_d|=5$ and ratio 3:1, one image goes to the test set; with $|S_d|=2$ and ratio 3:1, no image goes to the test set. The training set is the complement of the test set. We build splits for ratios from 1:1 to 9:1, repeat each with three random seeds, and report means and standard deviations of Top-1, Top-10, Top-50, and Top-100 accuracy. The resulting per-ratio split sizes are given in Table~\ref{tab:s1}. This construction is what makes the evaluation fair: by requiring that every test disorder has been seen with non-null features during training, it prevents the model from being penalized for disorders it could never have learned a textual signal for, and it isolates the effect of the image and text modalities from the confound of missing data.

\begin{table}[t]
\caption{Summary of optimal train-test splits for GMDB (v1.0.9) at different ratios.}
\label{tab:s1}
\centering
\begin{tabular}{lrrr}
\toprule
\textbf{Ratio} & \textbf{Train images} & \textbf{Test images} & \textbf{Test disorders}\\
\midrule
1:1 & 7632 & 2132 & 311\\
2:1 & 8407 & 1357 & 252\\
3:1 & 8795 & 969 & 227\\
4:1 & 9019 & 745 & 195\\
5:1 & 9159 & 605 & 175\\
6:1 & 9268 & 496 & 157\\
7:1 & 9358 & 406 & 137\\
8:1 & 9399 & 365 & 132\\
9:1 & 9453 & 311 & 123\\
\bottomrule
\end{tabular}
\end{table}

\subsection{Evaluation Metrics}
We report Top-N accuracy, defined as the fraction of test cases for which the true disease label appears within the top $N$ ranked predictions, for $N \in \{1, 10, 50, 100\}$. In a clinical prioritization setting these values answer complementary questions: Top-1 measures how often the single best guess is correct, while Top-10 to Top-100 measure how often the correct diagnosis is surfaced within a shortlist that a clinician or a downstream sequencing reinterpretation pipeline could feasibly review. Because the candidate space contains 528 disorders, even Top-50 represents a substantial narrowing of the differential.

\subsection{Equity and Clustering Analyses}
For the equity analysis we compare two training regimes at a matched size. One regime uses only patients of European ancestry; the other uses patients of all ancestries, with the number of European patients reduced by 72\% per disease so that the two training sets are comparable. In both regimes the test set is restricted to patients whose diagnostic disease already appears in the training set. We also vary the text component among three settings, namely no text, demographics only, and demographics combined with HPO terms, to separate the contribution of demographic context from that of clinical phenotypes. For the clustering analysis we apply two-component UMAP to the penultimate-layer logits, which form an $n \times 528$ matrix over the $n$ test samples, using $\mathrm{n\_neighbors}=7$, $\mathrm{min\_dist}=0.1$, and two components. The equity experiments start from an optimal 4:1 split repeated over three random seeds. After reducing the European patients by 72\% per disease, the all-ancestry and European-only training sets are closely matched in size, on average 5{,}375.7 versus 5{,}360 images, so that any accuracy difference reflects ancestry composition rather than training-set size; the per-ancestry training and test sizes for both regimes are listed in Table~\ref{tab:s3}.

%==============================================================
\section{Experiments and Results}
%==============================================================
\subsection{GestaltMML Classifies Rare Diseases in GMDB}
\begin{figure}[t]
\centering
\includegraphics[width=\columnwidth]{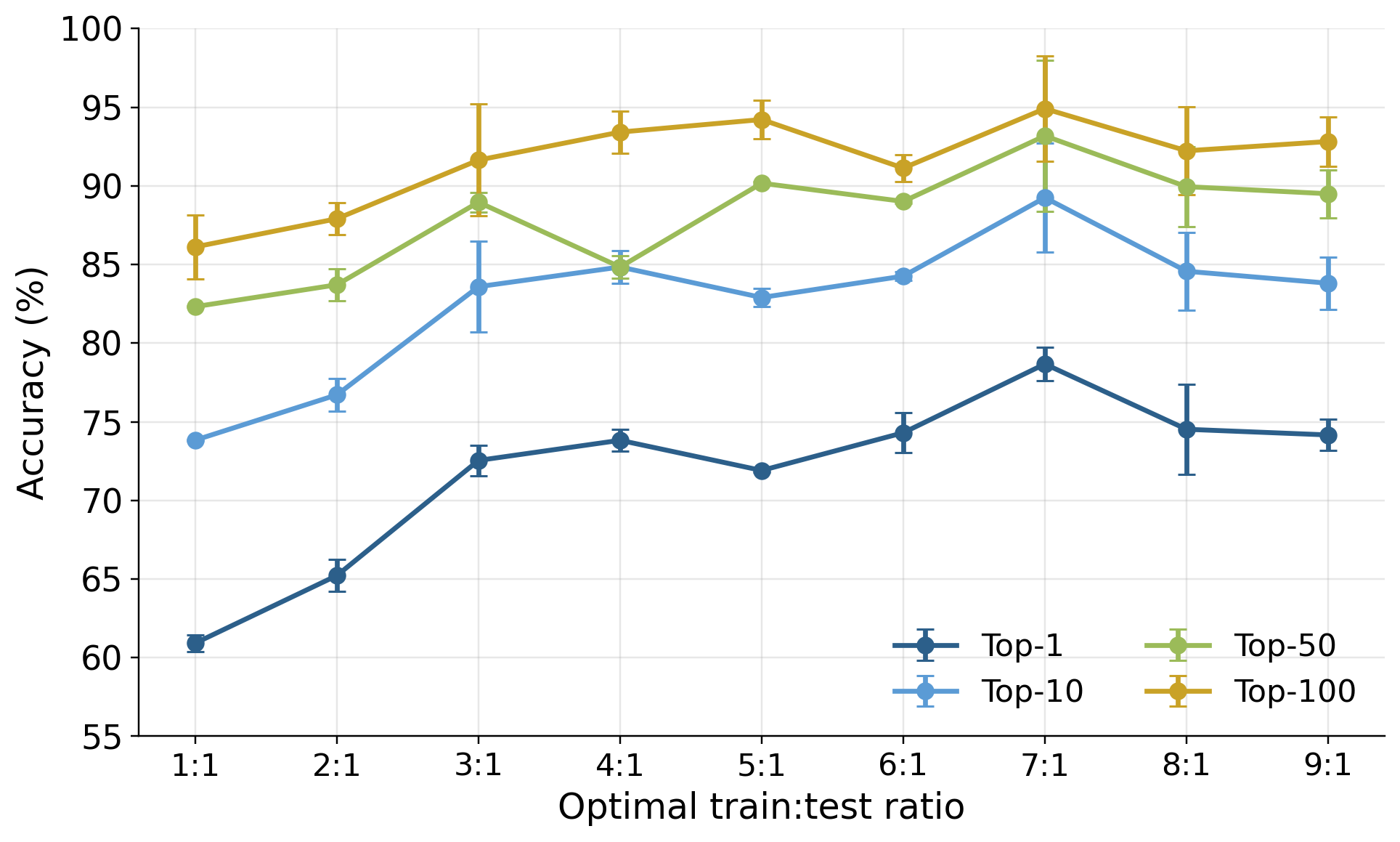}
\caption{GestaltMML accuracy as a function of the optimal train-test ratio (means with standard deviation over three seeds), corresponding to Table~\ref{tab:s2} in Appendix~\ref{app:split}. Accuracy rises sharply up to a 3:1 ratio and then plateaus.}
\label{fig:ratio}
\end{figure}

Top-N accuracy measures how often the true disease label appears among the top-N predictions, whether from image information alone or from combined image and textual information. With an optimal train-test ratio of 3:1, GestaltMML reaches mean accuracies of 72.54\% for Top-1, 83.59\% for Top-10, 88.96\% for Top-50, and 91.64\% for Top-100. The fact that Top-1 already exceeds 70\% with only three training images per held-out image is notable for a 528-class problem, and the further rise toward Top-100 shows that the correct diagnosis is almost always surfaced within a short candidate list. As Fig.~\ref{fig:ratio} shows (with split sizes in Table~\ref{tab:s1} and full numbers in Table~\ref{tab:s2}), accuracy rises sharply from the 1:1 to the 3:1 ratio and then largely plateaus, fluctuating between roughly 72\% and 79\% Top-1 at higher ratios, since each disorder is then represented by more training images while the test set becomes smaller. The plateau is informative for deployment: beyond a 3:1 ratio, adding more training images yields diminishing returns, and the apparent peak at 7:1 (78.67\% Top-1) coincides with a smaller and easier test set of only 406 images spanning 137 disorders (Table~\ref{tab:s1}). The higher-ratio numbers therefore come with larger standard deviations and should be read as a plateau rather than as strictly better operating points. Even at the modest 3:1 ratio, however, the model is already strong when images, demographics, and clinical features are all available, which is the realistic situation when a clinician records a patient's phenotype. We acknowledge two caveats. First, 528 disorders are a small number compared with the thousands of known rare diseases. Second, facial databases such as GMDB carry ascertainment biases, since only disorders with characteristic dysmorphic features tend to be documented. For diseases without machine-recognizable facial features the image branch may contribute little, but the demographic and clinical phenotype information in GestaltMML can still help prioritize candidate diagnoses, which is one reason multimodal inputs are valuable.

\subsection{Modality Importance and Comparison with the Image-Only Baseline}
Almost all existing literature has centered on facial images alone. We selected the up-to-date ensembled image model \cite{r13} as our benchmark and adopted the same train-test splits. Following that work, we separated GMDB into GMDB-frequent and GMDB-rare. For GMDB-frequent we used 7{,}755 images for training and 792 for testing; for GMDB-rare we used the same ten-fold cross validation, with on average 856.9 images for training and 360.1 for testing. Table~\ref{tab:main} reports the comparison among the full model, the two modality-masked variants, and the image-only baseline.

GestaltMML achieves the highest Top-1 and Top-10 accuracy on GMDB-frequent and improves on the ensembled image model, the strongest published image-only baseline, across all four reported metrics. On GMDB-rare, GestaltMML and GestaltLT are essentially tied: GestaltLT is 0.02 percentage points higher at Top-1 (24.16\% versus 24.14\%), while both reach 41.38\% at Top-10. The masked configurations are informative about why. GestaltViT, which sees images only, performs worst by a wide margin. This is the expected behavior of ViLT's linear patch embedding: with no convolutional or pretrained facial feature extractor, the image branch must learn facial structure from scratch, and under the limited data available for rare diseases it cannot match a dedicated face network. This is consistent with the known scaling behavior of Transformer vision models, which require large training sets to reach their potential \cite{r37}, and it implies that GestaltViT would improve substantially with more images. GestaltLT, which sees text only, is the opposite case: it remains close to the full model and clearly outperforms both GestaltViT and the ensembled image model. Two factors explain its strength. The clinical phenotype text is highly discriminative for many syndromes, and the OMIM-based augmentation supplies a meaningful textual signal even for images that lack curated HPO terms. Taken together, these results show that the text modality carries most of the predictive power. The image branch provides a measurable gain on GMDB-frequent, whereas no additional gain is observed on GMDB-rare. It is also worth emphasizing that GestaltMML is the only model in the comparison that is completely free of convolution, in contrast to the image-only systems.

The per-ancestry results in Appendix~\ref{app:equity} let us decompose this contribution at inference time. With the all-ancestry model, image-only inference yields very low Top-1 accuracy across groups, for example 1.79\% for European patients and 0.00\% for several minority groups (Table~\ref{tab:s6}). Adding demographic attributes gives a small but consistent lift, to 4.02\% for European and 2.86\% for Middle-East/West Asian patients (Table~\ref{tab:s5}), but the decisive gain comes from clinical text: full multimodal inference reaches 52.68\% for European, 65.11\% for Middle-East/West Asian, and 76.82\% for the Unknown group (Table~\ref{tab:s4}). This ordering quantifies what the GestaltLT and GestaltViT comparison already suggested, namely that in the present data regime the textual phenotype is the dominant signal, demographics contribute a smaller but reliable improvement, and the weak linear image branch contributes little on its own. It also explains why the model degrades gracefully when text is missing rather than failing outright, since the demographic prefix alone still carries some signal.

\begin{table}[t]
\caption{Performance on GMDB after fine-tuning (\%). Underlined variants are evaluated on the subset with the available test modality. Ensembled is the state-of-the-art iResNet and ArcFace image ensemble \cite{r13}. The best result in each column is shown in bold; tied results are both bolded.}
\label{tab:main}
\centering
\renewcommand{\arraystretch}{1.2}
\begin{tabular}{lcccc}
\toprule
& \multicolumn{2}{c}{\textbf{GMDB-frequent}} & \multicolumn{2}{c}{\textbf{GMDB-rare}}\\
\cmidrule(lr){2-3}\cmidrule(lr){4-5}
\textbf{Model} & \textbf{Top-1} & \textbf{Top-10} & \textbf{Top-1} & \textbf{Top-10}\\
\midrule
GestaltMML               & \textbf{50.14} & \textbf{75.50} & 24.14 & \textbf{41.38}\\
\underline{GestaltLT}    & 46.40 & 74.93 & \textbf{24.16} & \textbf{41.38}\\
\underline{GestaltViT}   & 18.16 & 44.67 & \phantom{0}6.97 & 18.12\\
Ensembled (image-only)   & 43.02 & 72.44 & 19.77 & 38.57\\
\bottomrule
\end{tabular}
\end{table}

\subsection{Diagnostic Equity for Under-Represented Groups}
A practical model must work not only on the majority population but also on patients from under-represented ancestries, who are precisely the groups most affected by biases in data collection and analysis. GMDB includes patients from many such groups, including Middle-East/West Asian, American-Native, South-East Asian, North African, African American, American-Latin/Hispanic, East Asian, Asian Others, South Asian, Sub-Saharan, and African, but they are heavily outnumbered by patients of European ancestry. We studied two questions. First, we asked which part of the text component drives accuracy by comparing three settings, no text, demographics only, and demographics combined with HPO terms. Clinical text has the largest effect, and demographic context provides an additional gain that is most pronounced for minority patients, who benefit when the model is told their age, sex, and ancestry rather than having to infer everything from a face that is under-represented in training. Second, we asked whether diversifying the training data helps. As shown in Fig.~\ref{fig:equity} for Top-1 and Top-10 accuracy, with the remaining metrics in Appendix~\ref{app:equity}, training GestaltMML on patients of all ancestries rather than on European patients only, at a matched training-set size, improves accuracy across under-represented groups with only rare exceptions. Taken together, these results indicate that incorporating multimodal context and diversifying the training set partially offset the bias introduced during data collection and lead to a fairer diagnostic procedure \cite{r81}. The magnitude of this effect is large for the most under-represented groups. Under full multimodal inference, moving from European-only to all-ancestry training raises Top-1 accuracy from 24.36\% to 65.11\% for Middle-East/West Asian patients, from 24.23\% to 76.82\% for the Unknown group, and from 22.11\% to 72.31\% for East Asian patients (Tables~\ref{tab:s7} and~\ref{tab:s4}). The European group is the only one that declines, from 72.06\% to 52.68\%, which is expected because its training data were reduced by 72\% to match the combined-cohort size. The net effect is a substantial narrowing of the accuracy gap across ancestries rather than a uniform gain, which is the more meaningful outcome for equitable deployment. Complete per-ethnicity results, with standard deviations, are given in Appendix~\ref{app:equity}.

\begin{figure}[t]
\centering
\includegraphics[width=\columnwidth]{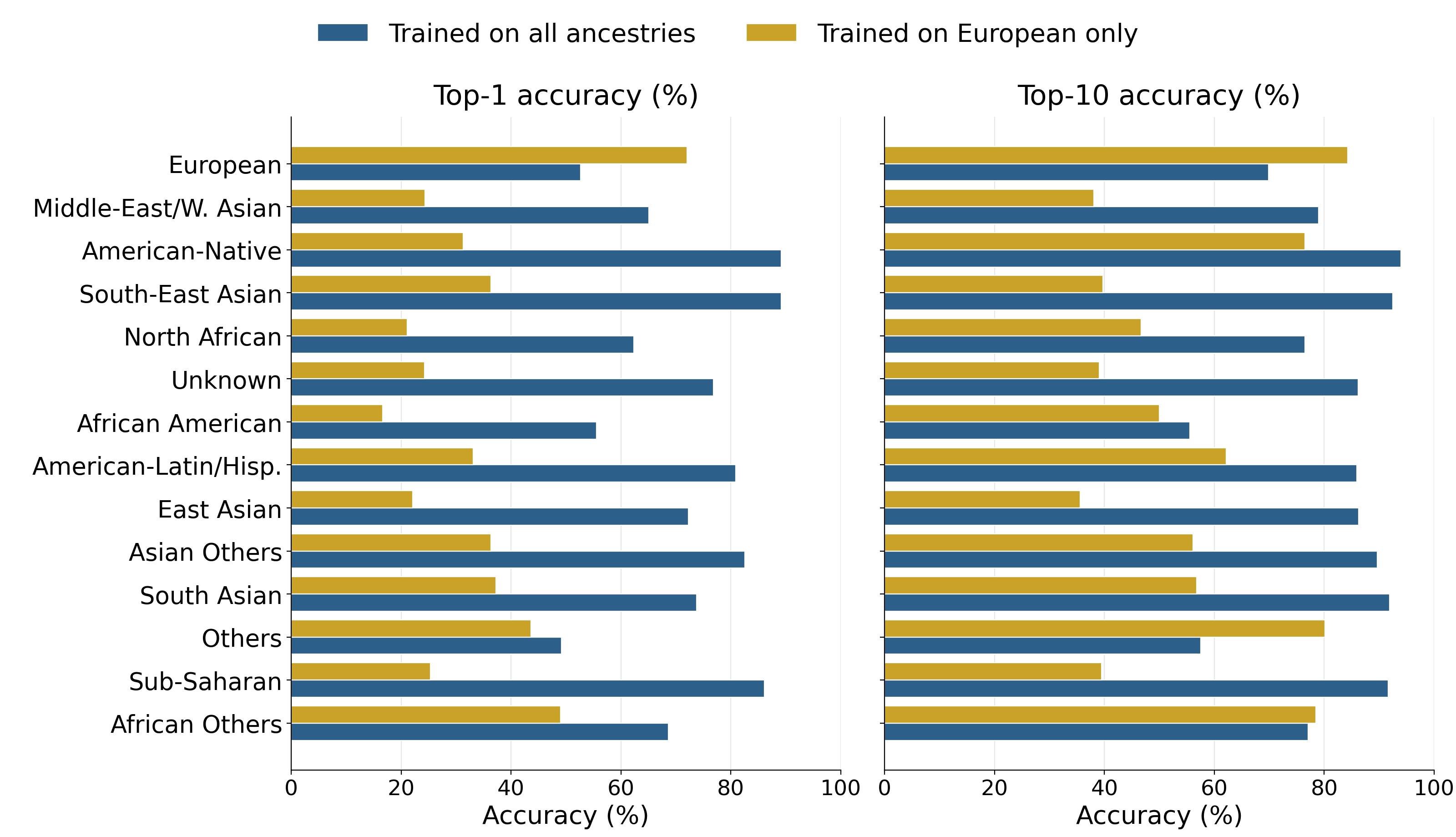}
\caption{Diagnostic equity. Top-1 and Top-10 accuracy of GestaltMML across ancestry groups under full multimodal inference, when trained on patients of all ancestries versus on European patients only, using matched training-set sizes. Training on all ancestries improves accuracy for nearly every under-represented group; the European group is the exception because its training data were reduced to match sizes. Per-ancestry results for all metrics and inference settings are in Appendix~\ref{app:equity} (Tables~\ref{tab:s4}--\ref{tab:s9}).}
\label{fig:equity}
\end{figure}

\subsection{External Validation}
We evaluate five external cohorts to assess robustness to potential bias in GMDB. Here we use GestaltMML trained at a 9:1 optimal ratio, except for NAA10, where we use a model trained on GMDB v1.0.3 at a 4:1 ratio to avoid train-test overlap. We also compare with the ensembled image model. Table~\ref{tab:external} reports Top-1 to Top-100 accuracy for every cohort and modality.

\textbf{Beckwith-Wiedemann syndrome (BWS).}
BWS is one of the most common overgrowth syndromes \cite{r47,r48,r49,r50,r51}. It involves molecular aberrations within a cluster of imprinted genes on chromosome 11p15, with loss of methylation at imprinting control region 2 (IC2) in about 50\% of patients and paternal uniparental isodisomy (pUPD11) in about 20\% \cite{r53}. Key signs of overgrowth such as macrosomia, organomegaly, and hemihypertrophy may not be represented in frontal facial images \cite{r52,r54,r55}, so image-based models can fall short. We collected two in-house BWS cohorts at our institute, one IC2 and one pUPD11. Clinical phenotypic descriptions are highly valuable, and GestaltMML reaches 100\% Top-1 accuracy with this information, while accuracy drops sharply with images alone, and the ensembled image model also struggles. The appendix figure (Fig.~\ref{fig:external}A) shows an example where images alone rank BWS sixth, but adding demographics and clinical text moves it to first.

\textbf{Sotos syndrome.}
Sotos syndrome is characterized by excess growth in early life, with greater height, weight, and head size, and frequent delays in motor, cognitive, and social development \cite{r56,r57,r58,r59}. Because many features cannot be represented in facial photos, this syndrome highlights the value of clinical text. From 23 patients at our institute, multimodal data greatly improves accuracy (Fig.~\ref{fig:external}B), where image information alone ranks Sotos 288th but adding demographics and clinical text moves it to first. GestaltMML most often misdiagnoses Sotos as Marshall-Smith syndrome \cite{r60}, which shares distinctive facial traits.

\textbf{NAA10-related neurodevelopmental syndrome.}
NAA10-related syndrome \cite{r61} is X-linked with a broad spectrum, ranging from a severe and often lethal cardiac phenotype in males \cite{r62} to intellectual disability in both sexes \cite{r63}. We collected 68 subjects from a collaborating institute, which are not in GMDB v1.0.3 but mostly appear in v1.0.9, so results here use the v1.0.3 model. Multimodal information improves accuracy in most cases, but in some instances text degrades the result, because the text of these patients resembles that of other neurodevelopmental labels in GMDB. Demographic information can also introduce bias, since most training patients are under five years old and of European ancestry, and a patient outside this range can yield unstable predictions.

\textbf{Cornelia de Lange syndrome (CdLS).}
CdLS is a genetically heterogeneous multiple malformation syndrome with growth restriction, upper-limb differences, hypertrichosis, long eyelashes, thick eyebrows, and other distinctive findings \cite{r64,r65,r66}. From 19 patients at our institute, CdLS diagnosis favored image-based over text-based analyses, and GestaltMML using only facial images surpassed multimodal inference, so the ensembled image model also reaches high accuracy. This reflects the distinctive facial features of CdLS, while text such as global developmental delay and feeding difficulties blurs distinctions with other neurodevelopmental syndromes.

\textbf{KBG syndrome.}
KBG syndrome is an extremely rare, pan-ethnic, autosomal dominant disorder with macrodontia, post-natal short stature, skeletal anomalies, abnormal hair implantation, and developmental delays \cite{r67,r68,r69,r70,r71}.From 18 samples at a collaborating institution, image-based models again outperformed both multimodal and text-only models. As with CdLS, common HPO terms such as global developmental delay overlap with many other disorders, and the older age of these patients relative to the training set introduced instability when age was encoded as text.

\textbf{When fusion helps.}
The five cohorts fall into two regimes that together yield a practical guideline. Fusion is decisive when the discriminating signal lives in non-facial phenotypes that a frontal photograph cannot show, as in the overgrowth syndromes BWS and Sotos, where adding text raises the correct diagnosis from a poor rank to the top. Fusion is counterproductive when the face is highly diagnostic and the recorded HPO terms are generic, as in CdLS and KBG, where terms such as global developmental delay and intellectual disability are shared by many neurodevelopmental disorders and pull the prediction toward the wrong label. NAA10 is the hardest case for every method and additionally illustrates a demographic failure mode, in which a test population that is older than the young, European-dominated training set destabilizes the text-encoded age signal. The actionable rule for deployment is therefore to prefer multimodal inference for facially subtle syndromes, to fall back to image-only inference when distinctive dysmorphology is present and the text is non-specific, and to interpret demographic-conditioned predictions cautiously when a patient lies outside the training distribution.

\begin{table}[t]
\caption{External validation accuracy (\%). Images+Texts is multimodal inference; Ensembled is the image-only ensemble \cite{r13}. Best per cohort and metric in bold. Cohort sizes (external $n$ / GMDB images $g$): Sotos 23/126, BWS-IC2 10/26, BWS-pUPD11 11/26, NAA10 68/15, CdLS 19/382, KBG 18/167.}
\label{tab:external}
\centering
\footnotesize
\renewcommand{\arraystretch}{1.12}
\setlength{\tabcolsep}{3pt}
\begin{tabular}{llrrrr}
\toprule
\textbf{Cohort} & \textbf{Modality} & \textbf{Top-1} & \textbf{Top-10} & \textbf{Top-50} & \textbf{Top-100}\\
\midrule
\multirow{4}{*}{Sotos} & Images+Texts & \textbf{73.91} & \textbf{86.96} & \textbf{95.65} & \textbf{95.65}\\
 & Texts & 69.57 & 82.61 & \textbf{95.65} & \textbf{95.65}\\
 & Images & 0.00 & 4.34 & 17.39 & 30.43\\
 & Ensembled & 41.93 & 61.29 & 93.55 & 95.16\\
\midrule
\multirow{4}{*}{BWS-IC2} & Images+Texts & \textbf{100.00} & \textbf{100.00} & \textbf{100.00} & \textbf{100.00}\\
 & Texts & 90.00 & \textbf{100.00} & \textbf{100.00} & \textbf{100.00}\\
 & Images & 10.00 & 40.00 & 70.00 & 90.00\\
 & Ensembled & 20.00 & 60.00 & 90.00 & \textbf{100.00}\\
\midrule
\multirow{4}{*}{BWS-pUPD11} & Images+Texts & \textbf{100.00} & \textbf{100.00} & \textbf{100.00} & \textbf{100.00}\\
 & Texts & \textbf{100.00} & \textbf{100.00} & \textbf{100.00} & \textbf{100.00}\\
 & Images & 0.00 & 27.27 & 45.45 & 54.54\\
 & Ensembled & 8.33 & 66.67 & 91.67 & \textbf{100.00}\\
\midrule
\multirow{4}{*}{NAA10} & Images+Texts & 4.41 & 32.35 & 66.18 & \textbf{82.35}\\
 & Texts & 5.88 & \textbf{38.23} & \textbf{72.06} & \textbf{82.35}\\
 & Images & 0.00 & 1.47 & 23.53 & 41.18\\
 & Ensembled & \textbf{7.35} & 16.18 & 44.12 & 66.18\\
\midrule
\multirow{4}{*}{CdLS} & Images+Texts & 21.05 & 73.68 & 84.21 & 94.74\\
 & Texts & 0.00 & 47.36 & 84.21 & 89.47\\
 & Images & 52.63 & 68.42 & 84.21 & 94.74\\
 & Ensembled & \textbf{76.67} & \textbf{86.67} & \textbf{100.00} & \textbf{100.00}\\
\midrule
\multirow{4}{*}{KBG} & Images+Texts & 44.44 & 83.33 & 88.89 & 88.89\\
 & Texts & 38.89 & 72.22 & 88.89 & 88.89\\
 & Images & 55.56 & 66.67 & 83.33 & 88.89\\
 & Ensembled & \textbf{94.44} & \textbf{94.44} & \textbf{100.00} & \textbf{100.00}\\
\bottomrule
\end{tabular}
\end{table}

\subsection{Clustering of Clinically Similar Diseases}
\begin{figure}[t]
\centering
\includegraphics[width=\columnwidth]{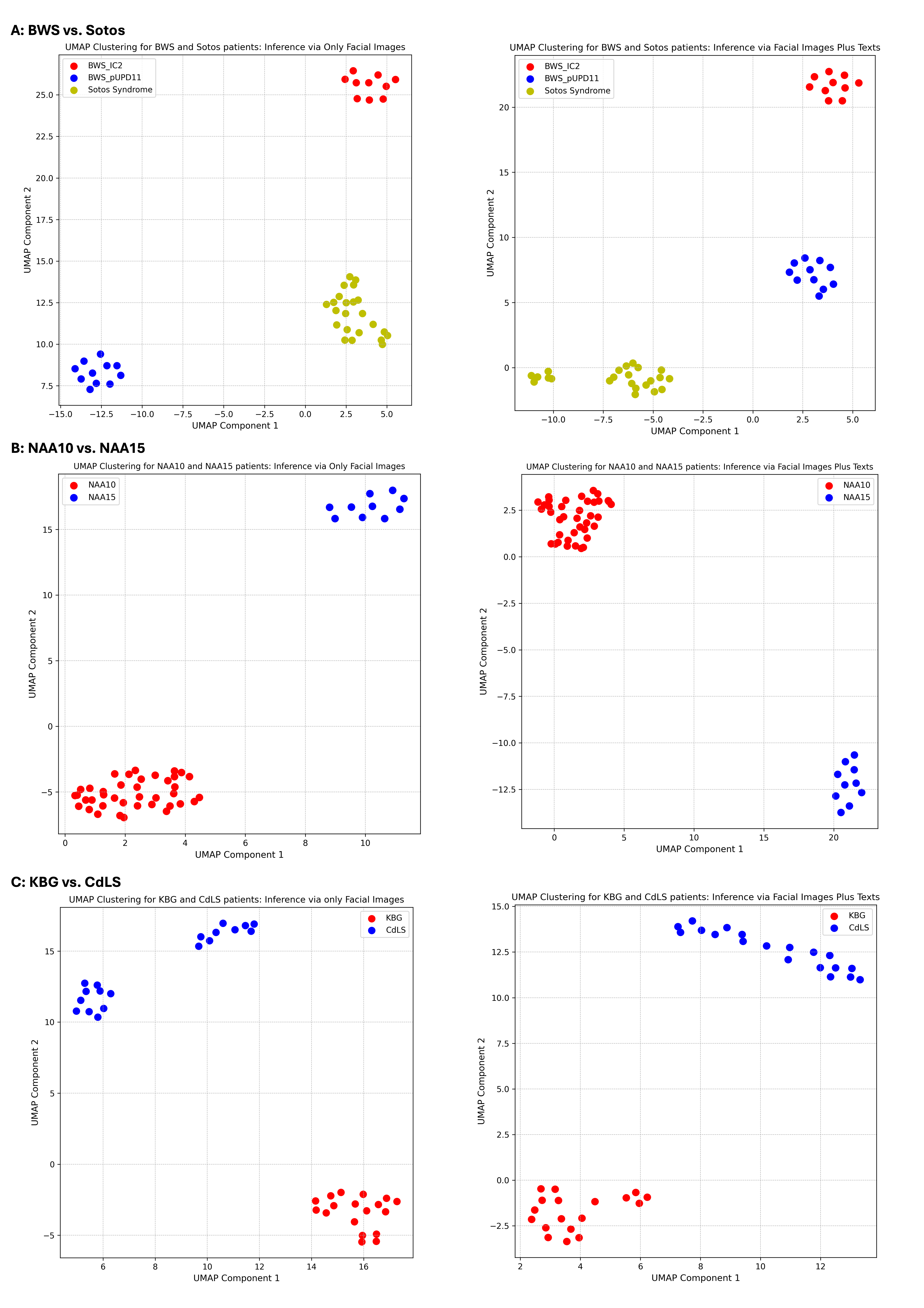}
\caption{UMAP clustering across three comparative sets. A: BWS cohorts (IC2 and pUPD11) with Sotos patients. B: NAA10 and NAA15 patients within GMDB (v1.0.9). C: KBG and CdLS patients. In each panel, the left plot uses facial images only and the right plot uses facial images plus texts.}
\label{fig:umap}
\end{figure}

Beyond classification, we asked whether the representation learned by GestaltMML organizes clinically related disorders in a meaningful way, a form of model interpretability \cite{r80}. We applied two-component UMAP to the penultimate-layer logits for three comparative sets that are known to be diagnostically confusable: BWS versus Sotos, NAA10 versus NAA15-related syndromes, and KBG versus CdLS (Fig.~\ref{fig:umap}). In every case the model separates the pair. The BWS versus Sotos comparison is the most striking, because both are overgrowth syndromes with overlapping presentations, yet the model not only distinguishes the two diseases but also separates the two molecular subtypes of BWS, IC2 and pUPD11, which is a finer distinction than the disease label itself. The NAA10 versus NAA15 comparison (GMDB v1.0.9) also separates cleanly despite their substantial phenotypic overlap. The KBG versus CdLS comparison also separates cleanly, though among CdLS patients the image-only embedding split by photographic background color rather than face, suggesting that normalizing background is a simple way to sharpen the image branch. As with GestaltMatcher and the ensembled image model, this clustering ability lets the model flag novel or previously unrecognized disorders without altering the classification layer or retraining, which is valuable for a field in which new syndromes are described every year.

%==============================================================
\section{Conclusion}
%==============================================================
We introduced GestaltMML, a convolution-free Transformer that fuses frontal facial photographs, demographic information, and clinical text to narrow the differential diagnosis of rare genetic diseases. On the GestaltMatcher Database it improves on the state-of-the-art image-only ensembled model, and modality masking shows that the clinical phenotype carries most of the predictive signal while the image branch still contributes; an OMIM-based augmentation supplies a textual signal even when curated HPO terms are missing; diversifying the training cohort and adding demographic context narrow the diagnostic-accuracy gap for under-represented ancestries; and across five external cohorts a simple deployment rule emerges, namely to prefer multimodal inference for facially subtle syndromes such as BWS and Sotos and to fall back to image-only inference when dysmorphology is distinctive and the recorded HPO terms are generic, as in CdLS and KBG. The principal limitation is the linear ViT-style image embedding \cite{r37}, which underperforms the text branch under scarce data; a stronger image encoder pretrained on patient faces, together with end-to-end phenotype extraction from free text \cite{r73}, is a promising direction. In combination with sequencing data, GestaltMML can support the interpretation and periodic reinterpretation of variants, ultimately helping to shorten the diagnostic odyssey.

% \section*{Acknowledgment}
% Acknowledgments, including funding sources and individual contributors, are omitted in this anonymized version to preserve double-blind review and will be restored in the camera-ready version.

%==============================================================
\clearpage
\newpage
\appendices
%==============================================================
% \section{Ethics, Consent, and Photos}
% \label{app:ethics}
% The study was approved by the Institutional Review Board of the Children's Hospital of Philadelphia (IRB 18-015712). The GMDB (v1.0.9) database was obtained from https://db.gestaltmatcher.org/. BWS and Sotos images were collected under CHOP IRB 13-010658, and CdLS under CHOP IRB 16-013231. For NAA10-related and KBG syndromes, oral and written consent were obtained under protocol \#7659 for the Jervis Clinic, approved by the New York State Psychiatric Institute and the Columbia University Department of Psychiatry IRB. Consent was obtained from all patients and legal guardians to analyze and, in some cases, publish the images.

\section{Ethics, Consent, and Photos}
\label{app:ethics}
The study was approved by the Institutional Review Boards of the participating institutions (approval numbers withheld for double-blind review). The GMDB (v1.0.9) database was obtained from the GestaltMatcher Database. The BWS, Sotos, and CdLS images were collected at our institution under the respective IRB approvals. For NAA10-related and KBG syndromes, oral and written consent were obtained at a collaborating institution under its approved protocol. Consent was obtained from all patients and legal guardians to analyze and, in some cases, publish the images.

\section{Dataset Distributions}
\label{app:dist}
\begin{figure}[!ht]
\centering
\includegraphics[width=\columnwidth]{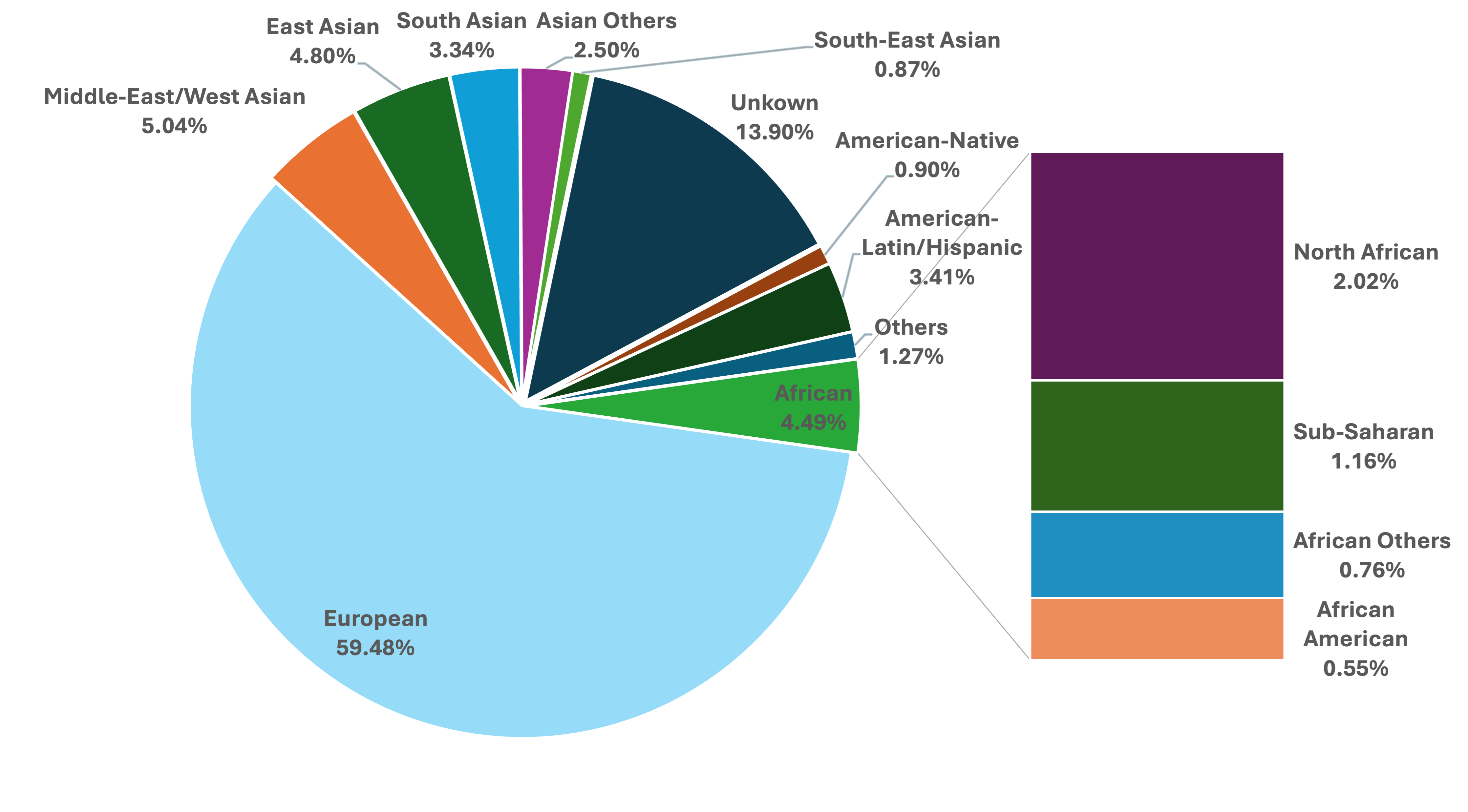}
\caption{Ancestry distribution within GMDB (v1.0.9).}
\label{fig:dist}
\end{figure}

Figure~\ref{fig:dist} shows the ancestry distribution within GMDB (v1.0.9), which is heavily skewed toward European ancestry (59.48\%). The sex distribution is approximately balanced, and the age distribution is uneven, with 64.90\% of patients under five years old.

\section{Optimal Train-Test Splits and Accuracy by Ratio}
\label{app:split}
Using the algorithm in Section~\ref{sec:optimal_splits}, we constructed optimal splits for ratios from 1:1 to 9:1, each repeated with three random seeds. Table~\ref{tab:s2} reports the accuracy at each ratio; the corresponding split sizes are given in the main text (Table~\ref{tab:s1}). The corresponding curve is shown in the main text (Fig.~\ref{fig:ratio}): accuracy rises sharply from the 1:1 to the 3:1 ratio and then largely plateaus at higher ratios, since each disorder is represented by more training images while the test set becomes smaller.

\begin{table}[h]
\caption{GestaltMML accuracy (\%, mean$\pm$SD over three seeds) at different optimal train-test ratios.}
\label{tab:s2}
\centering
\footnotesize
\setlength{\tabcolsep}{4pt}
\begin{tabular}{lrrrr}
\toprule
\textbf{Ratio} & \textbf{Top-1} & \textbf{Top-10} & \textbf{Top-50} & \textbf{Top-100}\\
\midrule
1:1 & 60.92$\pm$0.52 & 73.83$\pm$0.04 & 82.32$\pm$0.03 & 86.12$\pm$2.03\\
2:1 & 65.22$\pm$1.01 & 76.71$\pm$1.06 & 83.71$\pm$1.02 & 87.91$\pm$1.03\\
3:1 & 72.54$\pm$0.98 & 83.59$\pm$2.90 & 88.96$\pm$0.62 & 91.64$\pm$3.56\\
4:1 & 73.82$\pm$0.71 & 84.84$\pm$1.02 & 84.84$\pm$0.70 & 93.42$\pm$1.35\\
5:1 & 71.90$\pm$0.11 & 82.90$\pm$0.59 & 90.17$\pm$0.12 & 94.22$\pm$1.23\\
6:1 & 74.30$\pm$1.28 & 84.26$\pm$0.27 & 89.01$\pm$0.14 & 91.13$\pm$0.86\\
7:1 & 78.67$\pm$1.05 & 89.25$\pm$3.47 & 93.18$\pm$4.82 & 94.91$\pm$3.36\\
8:1 & 74.52$\pm$2.86 & 84.57$\pm$2.46 & 89.95$\pm$2.53 & 92.23$\pm$2.81\\
9:1 & 74.16$\pm$0.98 & 83.81$\pm$1.65 & 89.50$\pm$1.52 & 92.82$\pm$1.58\\
\bottomrule
\end{tabular}
\end{table}

\section{Diagnostic Equity: Detailed Per-Ancestry Results}
\label{app:equity}
For the equity experiments we begin from an optimal 4:1 split with three random seeds, using patients of European descent for the entire training set, then compare against a training set that includes patients of all ancestries with the same seeds. To keep the sizes comparable, we reduce the number of European patients by 72\% per disease. Table~\ref{tab:s3} lists the resulting training and test sizes. Tables~\ref{tab:s4}--\ref{tab:s9} give the complete per-ancestry accuracy for the two training regimes under three inference settings (images with demographics and text, images with demographics, and images only). The main-text equity figure (Fig.~\ref{fig:equity}) visualizes the full-modality rows of Tables~\ref{tab:s4} and~\ref{tab:s7}: training on all ancestries markedly improves accuracy for nearly every under-represented group, while the European group itself is lower under the matched-size all-ancestry regime because its training data were reduced.

\begin{table}[h]
\caption{Average training and test set sizes for the diagnostic-equity experiments, over three seeds.}
\label{tab:s3}
\centering
\footnotesize
\setlength{\tabcolsep}{4pt}
\begin{tabular}{lrrrr}
\toprule
& \multicolumn{2}{c}{\textbf{All ancestries}} & \multicolumn{2}{c}{\textbf{European only}}\\
\cmidrule(lr){2-3}\cmidrule(lr){4-5}
\textbf{Ancestry} & \textbf{Train} & \textbf{Test} & \textbf{Train} & \textbf{Test}\\
\midrule
European & 1712.3 & 452.3 & 5360 & 447.0\\
ME/W. Asian & 458.3 & 33.7 & 0 & 31.3\\
Am.-Native & 78.3 & 9.7 & 0 & 11.3\\
SE Asian & 77.3 & 7.7 & 0 & 8.0\\
N. African & 179.7 & 17.3 & 0 & 11.7\\
Unknown & 1238.3 & 118.7 & 0 & 112.3\\
Afr. American & 52.0 & 2.0 & 0 & 2.0\\
Am.-Latin/Hisp. & 313.0 & 20.0 & 0 & 18.3\\
East Asian & 444.3 & 24.7 & 0 & 25.3\\
Asian Others & 228.0 & 16.0 & 0 & 15.3\\
South Asian & 305.7 & 20.3 & 0 & 18.7\\
Others & 117.0 & 7.0 & 0 & 5.7\\
Sub-Saharan & 108.0 & 5.0 & 0 & 6.0\\
African Others & 63.3 & 10.7 & 0 & 7.0\\
\midrule
Total & 5375.67 & 745.0 & 5360 & 722.0\\
\bottomrule
\end{tabular}
\end{table}

\begin{table}[t]
\caption{Per-ancestry accuracy (\%) when trained on all ancestries, using images, demographics, and clinical.}
\label{tab:s4}
\centering
\footnotesize
\setlength{\tabcolsep}{2pt}
\begin{tabular}{lrrrr}
\toprule
\textbf{Ancestry} & \textbf{Top-1} & \textbf{Top-10} & \textbf{Top-50} & \textbf{Top-100}\\
\midrule
European & 52.68$\pm$1.24 & 69.85$\pm$1.35 & 79.65$\pm$1.33 & 84.53$\pm$0.43\\
ME/W. Asian & 65.11$\pm$10.53 & 78.96$\pm$12.38 & 84.91$\pm$10.60 & 92.99$\pm$4.70\\
Am.-Native & 89.18$\pm$9.57 & 93.94$\pm$4.50 & 93.94$\pm$4.50 & 96.97$\pm$3.25\\
SE Asian & 89.17$\pm$10.10 & 92.50$\pm$6.61 & 92.50$\pm$6.61 & 92.50$\pm$6.61\\
N. African & 62.32$\pm$16.40 & 76.47$\pm$13.16 & 81.87$\pm$9.05 & 89.35$\pm$7.27\\
Unknown & 76.82$\pm$6.00 & 86.16$\pm$3.04 & 91.08$\pm$3.40 & 94.11$\pm$2.14\\
Afr. American & 55.56$\pm$30.91 & 55.56$\pm$30.91 & 55.56$\pm$30.91 & 55.56$\pm$30.91\\
Am.-Latin/Hisp. & 80.89$\pm$4.86 & 85.94$\pm$6.18 & 88.86$\pm$3.65 & 88.86$\pm$3.65\\
East Asian & 72.31$\pm$8.79 & 86.28$\pm$3.39 & 90.37$\pm$3.06 & 91.56$\pm$4.74\\
Asian Others & 82.54$\pm$5.72 & 89.68$\pm$4.42 & 91.53$\pm$7.50 & 94.71$\pm$5.57\\
South Asian & 73.80$\pm$1.14 & 91.88$\pm$2.51 & 95.22$\pm$4.56 & 98.48$\pm$2.62\\
Others & 49.17$\pm$1.12 & 57.50$\pm$6.61 & 68.33$\pm$10.10 & 85.00$\pm$4.33\\
Sub-Saharan & 86.11$\pm$12.73 & 91.67$\pm$7.43 & 91.67$\pm$7.43 & 91.67$\pm$7.43\\
African Others & 68.63$\pm$1.75 & 77.09$\pm$9.32 & 89.26$\pm$11.13 & 96.67$\pm$3.22\\
\bottomrule
\end{tabular}
\end{table}

\begin{table}[t]
\caption{Per-ancestry accuracy (\%) when trained on all ancestries, using images and demographics.}
\label{tab:s5}
\centering
\footnotesize
\setlength{\tabcolsep}{2pt}
\begin{tabular}{lrrrr}
\toprule
\textbf{Ancestry} & \textbf{Top-1} & \textbf{Top-10} & \textbf{Top-50} & \textbf{Top-100}\\
\midrule
European & 4.02$\pm$1.74 & 20.09$\pm$3.82 & 35.04$\pm$4.82 & 45.09$\pm$4.09\\
ME/W. Asian & 2.86$\pm$1.71 & 25.71$\pm$3.74 & 45.71$\pm$5.11 & 57.14$\pm$1.70\\
Am.-Native & 9.09$\pm$2.23 & 72.72$\pm$6.23 & 81.82$\pm$5.24 & 90.91$\pm$5.24\\
SE Asian & 0.00$\pm$0.00 & 12.50$\pm$3.67 & 37.50$\pm$6.67 & 37.50$\pm$6.67\\
N. African & 0.00$\pm$0.00 & 0.00$\pm$0.00 & 26.67$\pm$1.43 & 53.33$\pm$1.19\\
Unknown & 3.13$\pm$1.53 & 23.44$\pm$1.91 & 40.63$\pm$5.22 & 48.44$\pm$3.22\\
Afr. American & 0.00$\pm$0.00 & 0.00$\pm$0.00 & 0.00$\pm$0.00 & 0.00$\pm$0.00\\
Am.-Latin/Hisp. & 9.52$\pm$1.43 & 42.86$\pm$1.22 & 57.14$\pm$1.96 & 66.67$\pm$1.71\\
East Asian & 0.00$\pm$0.00 & 10.71$\pm$4.11 & 42.86$\pm$6.52 & 57.14$\pm$4.86\\
Asian Others & 0.00$\pm$0.00 & 11.11$\pm$1.57 & 44.44$\pm$2.78 & 55.56$\pm$5.73\\
South Asian & 5.26$\pm$2.74 & 15.79$\pm$2.54 & 36.84$\pm$4.56 & 36.84$\pm$4.56\\
Others & 0.00$\pm$0.00 & 25.00$\pm$2.17 & 50.00$\pm$2.54 & 62.50$\pm$3.27\\
Sub-Saharan & 0.00$\pm$0.00 & 0.00$\pm$0.00 & 0.00$\pm$0.00 & 0.00$\pm$0.00\\
African Others & 0.00$\pm$0.00 & 0.00$\pm$0.00 & 11.11$\pm$3.51 & 11.11$\pm$3.51\\
\bottomrule
\end{tabular}
\end{table}

\begin{table}[t]
\caption{Per-ancestry accuracy (\%) when trained on all ancestries, using images only.}
\label{tab:s6}
\centering
\footnotesize
\setlength{\tabcolsep}{2pt}
\begin{tabular}{lrrrr}
\toprule
\textbf{Ancestry} & \textbf{Top-1} & \textbf{Top-10} & \textbf{Top-50} & \textbf{Top-100}\\
\midrule
European & 1.79$\pm$0.74 & 4.02$\pm$1.02 & 8.04$\pm$3.30 & 20.54$\pm$4.56\\
ME/W. Asian & 0.00$\pm$0.00 & 2.86$\pm$0.92 & 14.29$\pm$3.06 & 22.86$\pm$1.59\\
Am.-Native & 0.00$\pm$0.00 & 0.00$\pm$0.00 & 0.00$\pm$0.00 & 9.09$\pm$3.50\\
SE Asian & 0.00$\pm$0.00 & 12.50$\pm$3.83 & 25.00$\pm$1.37 & 25.00$\pm$1.37\\
N. African & 0.00$\pm$0.00 & 0.00$\pm$0.00 & 33.33$\pm$2.88 & 40.00$\pm$5.89\\
Unknown & 0.00$\pm$0.00 & 3.13$\pm$2.95 & 9.38$\pm$4.28 & 17.19$\pm$4.16\\
Afr. American & 0.00$\pm$0.00 & 0.00$\pm$0.00 & 0.00$\pm$0.00 & 0.00$\pm$0.00\\
Am.-Latin/Hisp. & 0.00$\pm$0.00 & 4.75$\pm$2.66 & 23.81$\pm$1.92 & 42.86$\pm$1.53\\
East Asian & 10.71$\pm$0.52 & 28.57$\pm$5.79 & 35.71$\pm$2.06 & 35.71$\pm$1.72\\
Asian Others & 0.00$\pm$0.00 & 0.00$\pm$0.00 & 11.11$\pm$2.24 & 22.22$\pm$2.91\\
South Asian & 0.00$\pm$0.00 & 0.00$\pm$0.00 & 5.26$\pm$3.28 & 10.53$\pm$3.95\\
Others & 0.00$\pm$0.00 & 0.00$\pm$0.00 & 12.50$\pm$3.53 & 25.00$\pm$2.47\\
Sub-Saharan & 0.00$\pm$0.00 & 0.00$\pm$0.00 & 25.00$\pm$5.89 & 25.00$\pm$5.89\\
African Others & 0.00$\pm$0.00 & 0.00$\pm$0.00 & 0.00$\pm$0.00 & 22.22$\pm$1.21\\
\bottomrule
\end{tabular}
\end{table}

\begin{table}[t]
\caption{Per-ancestry accuracy (\%) when trained on European patients only, using images, demographics, and clinical text.}
\label{tab:s7}
\centering
\footnotesize
\setlength{\tabcolsep}{2pt}
\begin{tabular}{lrrrr}
\toprule
\textbf{Ancestry} & \textbf{Top-1} & \textbf{Top-10} & \textbf{Top-50} & \textbf{Top-100}\\
\midrule
European & 72.06$\pm$2.51 & 84.28$\pm$2.96 & 89.79$\pm$2.21 & 92.25$\pm$1.04\\
ME/W. Asian & 24.36$\pm$3.21 & 38.08$\pm$9.68 & 56.25$\pm$7.57 & 66.64$\pm$9.98\\
Am.-Native & 31.31$\pm$3.35 & 76.51$\pm$4.73 & 85.35$\pm$13.75 & 85.35$\pm$13.75\\
SE Asian & 36.39$\pm$3.58 & 39.72$\pm$4.91 & 68.61$\pm$3.81 & 79.44$\pm$4.19\\
N. African & 21.11$\pm$1.70 & 46.67$\pm$1.55 & 62.22$\pm$3.85 & 84.44$\pm$9.62\\
Unknown & 24.23$\pm$3.00 & 39.05$\pm$3.06 & 54.82$\pm$4.34 & 65.27$\pm$3.33\\
Afr. American & 16.67$\pm$8.95 & 50.00$\pm$30.00 & 50.00$\pm$30.00 & 66.67$\pm$37.73\\
Am.-Latin/Hisp. & 33.10$\pm$3.93 & 62.14$\pm$4.34 & 68.49$\pm$4.02 & 80.71$\pm$6.50\\
East Asian & 22.11$\pm$6.51 & 35.58$\pm$1.97 & 47.25$\pm$4.02 & 56.17$\pm$6.50\\
Asian Others & 36.35$\pm$2.81 & 56.14$\pm$12.00 & 70.66$\pm$14.39 & 81.68$\pm$6.75\\
South Asian & 37.26$\pm$9.86 & 56.77$\pm$3.04 & 74.76$\pm$2.18 & 86.45$\pm$6.08\\
Others & 43.65$\pm$2.76 & 80.16$\pm$2.60 & 84.92$\pm$1.35 & 90.48$\pm$1.50\\
Sub-Saharan & 25.37$\pm$1.45 & 39.44$\pm$1.28 & 46.85$\pm$1.05 & 58.89$\pm$8.39\\
African Others & 48.98$\pm$11.92 & 78.43$\pm$8.76 & 85.09$\pm$6.46 & 88.43$\pm$11.14\\
\bottomrule
\end{tabular}
\end{table}

\begin{table}[t]
\caption{Per-ancestry accuracy (\%) when trained on European patients only, using images and demographics.}
\label{tab:s8}
\centering
\footnotesize
\setlength{\tabcolsep}{2pt}
\begin{tabular}{lrrrr}
\toprule
\textbf{Ancestry} & \textbf{Top-1} & \textbf{Top-10} & \textbf{Top-50} & \textbf{Top-100}\\
\midrule
European & 4.25$\pm$1.91 & 21.70$\pm$2.56 & 45.19$\pm$7.21 & 56.60$\pm$8.54\\
ME/W. Asian & 3.33$\pm$0.89 & 16.67$\pm$4.71 & 20.00$\pm$5.38 & 26.67$\pm$1.83\\
Am.-Native & 0.00$\pm$0.00 & 4.55$\pm$1.42 & 63.64$\pm$5.82 & 63.64$\pm$3.77\\
SE Asian & 25.00$\pm$4.12 & 37.50$\pm$2.65 & 62.50$\pm$2.12 & 62.50$\pm$7.07\\
N. African & 0.00$\pm$0.00 & 10.00$\pm$0.00 & 20.00$\pm$0.00 & 50.00$\pm$0.00\\
Unknown & 1.67$\pm$0.06 & 13.33$\pm$0.65 & 30.00$\pm$3.62 & 36.67$\pm$8.32\\
Afr. American & 0.00$\pm$0.00 & 0.00$\pm$0.00 & 50.00$\pm$0.00 & 50.00$\pm$0.00\\
Am.-Latin/Hisp. & 10.00$\pm$0.00 & 15.00$\pm$0.00 & 45.00$\pm$4.48 & 60.00$\pm$3.74\\
East Asian & 0.00$\pm$0.00 & 14.81$\pm$1.66 & 40.74$\pm$1.25 & 48.15$\pm$2.05\\
Asian Others & 0.00$\pm$0.00 & 11.11$\pm$1.12 & 22.22$\pm$4.49 & 22.22$\pm$5.10\\
South Asian & 5.26$\pm$1.53 & 15.79$\pm$3.90 & 52.63$\pm$4.28 & 57.89$\pm$5.95\\
Others & 33.33$\pm$0.00 & 50.00$\pm$0.00 & 66.67$\pm$3.53 & 66.67$\pm$3.53\\
Sub-Saharan & 0.00$\pm$0.00 & 0.00$\pm$0.00 & 0.00$\pm$0.00 & 0.00$\pm$0.00\\
African Others & 0.00$\pm$0.00 & 33.33$\pm$1.42 & 33.33$\pm$1.42 & 44.44$\pm$1.51\\
\bottomrule
\end{tabular}
\end{table}

\begin{table}[t]
\caption{Per-ancestry accuracy (\%) when trained on European patients only, using images only.}
\label{tab:s9}
\centering
\footnotesize
\setlength{\tabcolsep}{2pt}
\begin{tabular}{lrrrr}
\toprule
\textbf{Ancestry} & \textbf{Top-1} & \textbf{Top-10} & \textbf{Top-50} & \textbf{Top-100}\\
\midrule
European & 0.22$\pm$0.03 & 2.24$\pm$0.56 & 12.30$\pm$3.21 & 27.07$\pm$4.54\\
ME/W. Asian & 0.00$\pm$0.00 & 0.00$\pm$0.00 & 16.67$\pm$5.38 & 33.33$\pm$1.83\\
Am.-Native & 0.00$\pm$0.00 & 9.09$\pm$1.42 & 9.09$\pm$1.42 & 27.27$\pm$3.77\\
SE Asian & 0.00$\pm$0.00 & 0.00$\pm$0.00 & 25.00$\pm$2.12 & 25.00$\pm$2.12\\
N. African & 0.00$\pm$0.00 & 0.00$\pm$0.00 & 0.00$\pm$0.00 & 10.00$\pm$3.07\\
Unknown & 2.50$\pm$0.06 & 3.33$\pm$0.65 & 12.50$\pm$3.62 & 21.67$\pm$8.32\\
Afr. American & 0.00$\pm$0.00 & 0.00$\pm$0.00 & 0.00$\pm$0.00 & 0.00$\pm$0.00\\
Am.-Latin/Hisp. & 0.00$\pm$0.00 & 0.00$\pm$0.00 & 20.00$\pm$4.48 & 40.00$\pm$0.70\\
East Asian & 0.00$\pm$0.00 & 0.00$\pm$0.00 & 7.41$\pm$1.25 & 25.93$\pm$2.05\\
Asian Others & 11.11$\pm$1.12 & 11.11$\pm$1.12 & 11.11$\pm$1.12 & 44.44$\pm$10.10\\
South Asian & 0.00$\pm$0.00 & 0.00$\pm$0.00 & 21.05$\pm$4.28 & 26.31$\pm$5.95\\
Others & 0.00$\pm$0.00 & 0.00$\pm$0.00 & 50.00$\pm$3.53 & 50.00$\pm$3.53\\
Sub-Saharan & 0.00$\pm$0.00 & 0.00$\pm$0.00 & 0.00$\pm$0.00 & 25.00$\pm$5.89\\
African Others & 0.00$\pm$0.00 & 0.00$\pm$0.00 & 0.00$\pm$0.00 & 0.00$\pm$0.00\\
\bottomrule
\end{tabular}
\end{table}

\section{External Validation Example}
\label{app:cohorts}
\begin{figure}[h]
\centering
\includegraphics[width=\columnwidth]{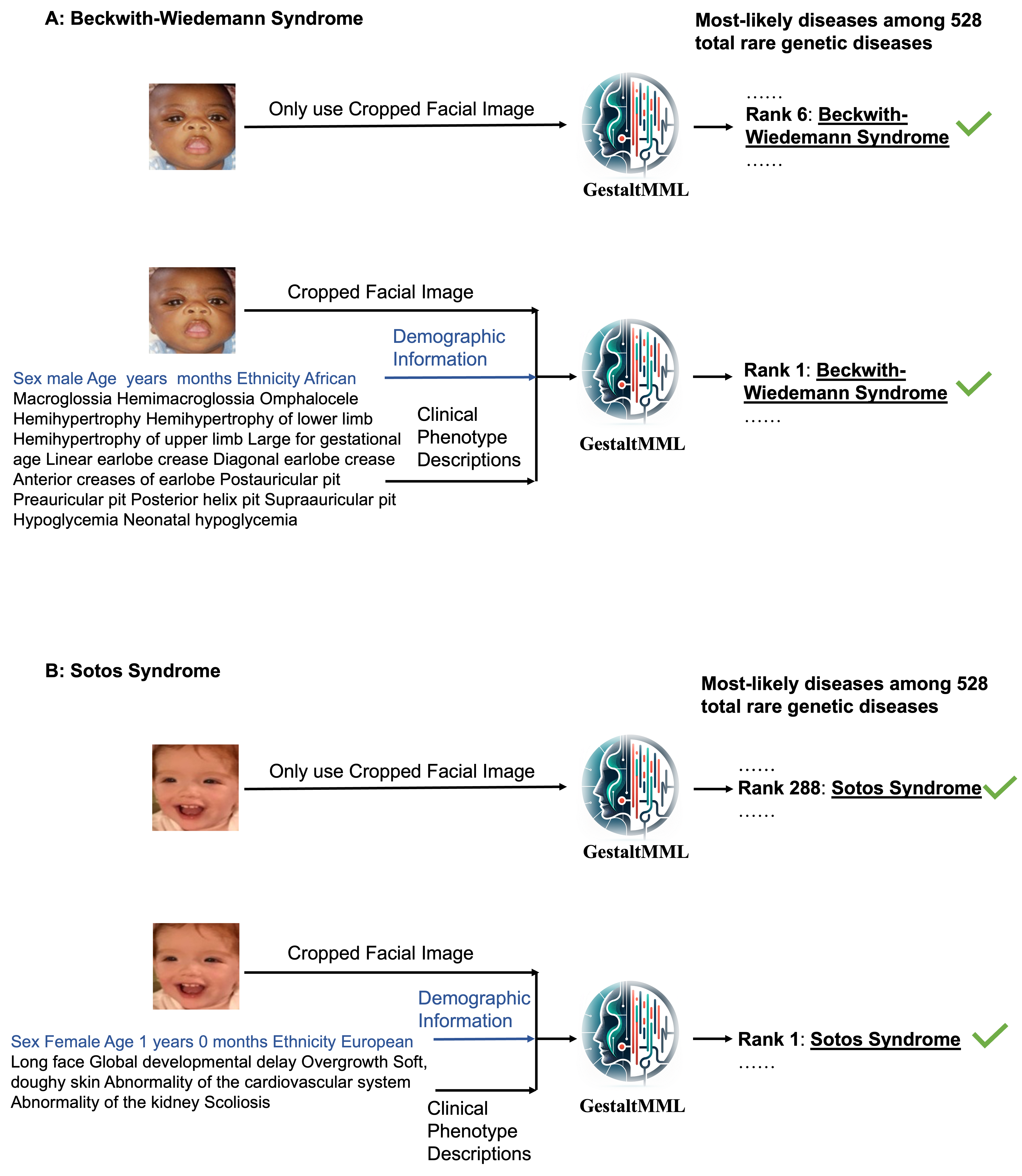}
\caption{Multimodal inference combining facial images, demographics, and clinical phenotype descriptions is more effective than facial images alone for BWS (A) and Sotos syndrome (B). Patient data were sourced from our institute with appropriate consent.}
\label{fig:external}
\end{figure}
Figure~\ref{fig:external} illustrates the value of multimodal fusion on two example patients, showing how adding demographic and clinical-text information moves the correct diagnosis from a low rank under image-only inference (rank six for the Beckwith-Wiedemann syndrome patient in A and rank 288 for the Sotos syndrome patient in B) to the top of the ranked list.

\section{NAA10 Per-Patient Ranking}
\label{app:naa10}
\begin{figure}[t]
\centering
\includegraphics[width=\columnwidth]{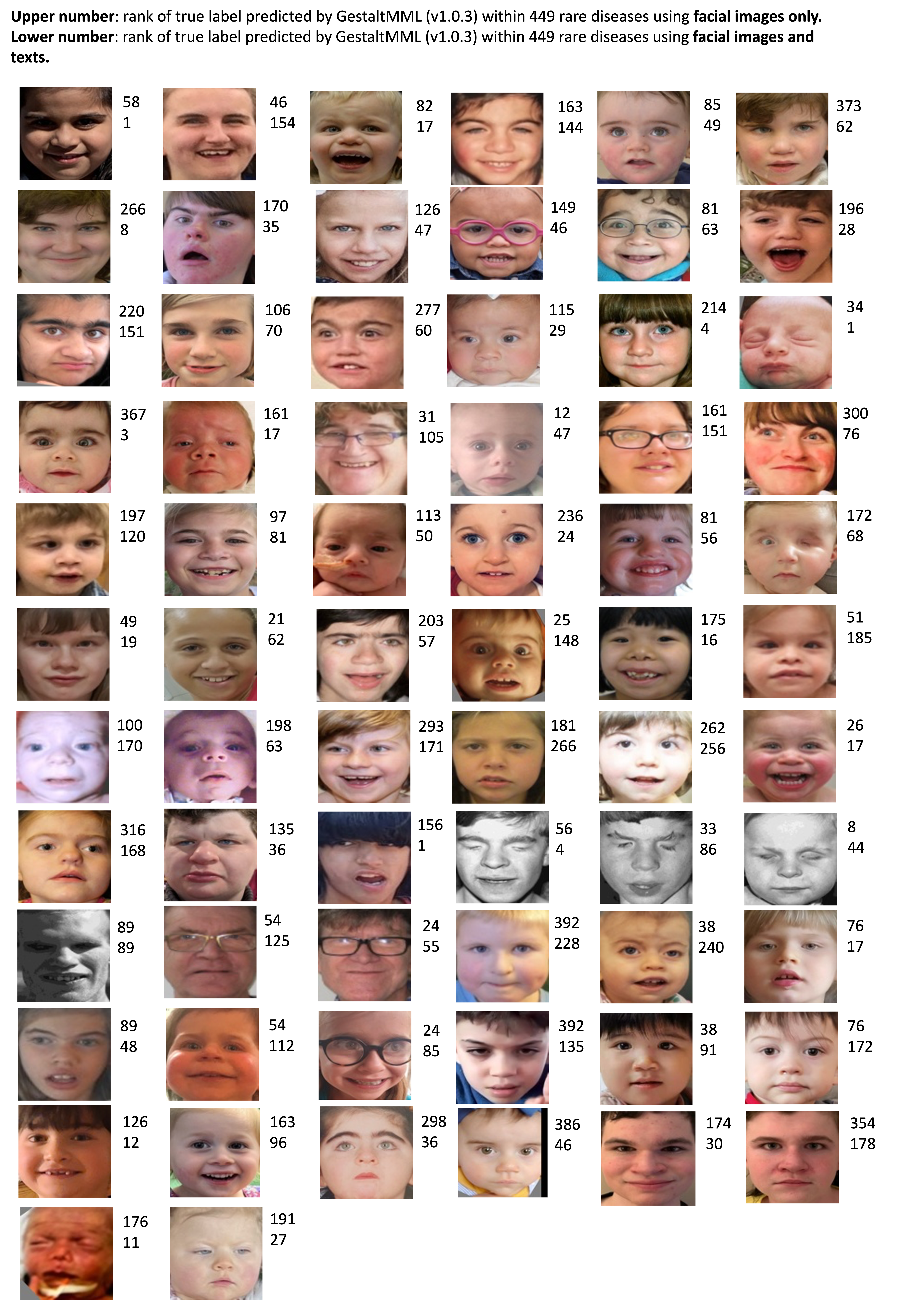}
\caption{Rank of the true label among 449 disease labels for 68 NAA10 patients from a collaborating institute and published literature, predicted by GestaltMML trained on GMDB v1.0.3 at a 4:1 ratio. Upper and lower numbers indicate the rank using facial images only and using combined information, respectively.}
\label{fig:naa10}
\end{figure}

Figure~\ref{fig:naa10} shows the rank of the true label among 449 disease labels for 68 NAA10 patients, comparing facial images only with combined information.

\section{Data and Code Availability}
\label{app:avail}
The GMDB (v1.0.9) database is available to researchers on application from the GestaltMatcher Database. All software tools and the computational workflow, provided as a Jupyter Notebook, will be released publicly upon publication. This study did not generate any new material. The authors declare no competing interests.

%==============================================================

\end{document}